\documentclass{article}

\usepackage{eus}
\usepackage[small]{caption2}
\usepackage{amsmath} 
\usepackage{amssymb}
\usepackage{epsfig}
\usepackage[latin1]{inputenc}
\usepackage[english]{babel}

\newcommand{\varP}{\gamma_i^2 \left( \sigma_{X_i}^2 + \alpha_i^2 \left( n-1 \right)
\right) + \sigma_{\delta_i}^2}

\title{
	Information-theoretic resolution of perceptual WSS watermarking
	of non i.i.d. Gaussian signals
}

\author{
	{\it Ga\"etan Le Guelvouit, St\'ephane Pateux and Christine Guillemot}\\
	IRISA/INRIA, Campus de Beaulieu, \\
	35042 Rennes Cedex, FRANCE\\
	Tel: +33 2 99 84 73 60; fax: +33 2 99 84 25 31\\
	{\tt e-mail: Gaetan.Le\_Guelvouit@irisa.fr}\\[2ex]
}

\begin{document}

\maketitle

\begin{abstract} \small
	The theoretical foundations of data hiding have been revealed by
	formulating the problem as message communication over a noisy channel.
	We revisit the problem in light of a more general characterization of
	the watermark channel and of weighted distortion measures. Considering
	spread spectrum based information hiding, we release the usual
	assumption of an i.i.d. cover signal. The game-theoretic resolution of
	the problem reveals a generalized characterization of optimum attacks.
	The paper then derives closed-form expressions for the different
	parameters exhibiting a practical embedding and extraction technique.
\end{abstract}

\section{INTRODUCTION}
	Information hiding refers to nearly invisible embedding of a message
	within a host signal. This paper focuses on the data hiding problem,
	assuming a blind and symmetric system. In the spirit of a
	communication problem one seeks the maximum rate of reliable
	transmission over any hiding and attack strategies. This rate is
	called the hiding capacity and depends on admissible distortion levels
	and on the watermark channel characterization.

	Watermarking is often regarded as a form of spread spectrum
	communication with various forms of channel characterizations. The
	perceptual sensitivity of the host signal is often taken into account
	for choosing embedding sites and
	strength~\cite{CoxKilianLeightonShamoon97,PodilchukZeng98}. The
	attacks are often modelled as the addition of White Gaussian noise
	(AWGN)~\cite{ServettoPodilchukRamchandran98,MoulinSullivan2000}, or as
	linear filtering plus additive
	noise~\cite{SuEggersGirod2001,MoulinIvanovic2001}. The authors
	in~\cite{SuGirod99} show that the optimum attack is obtained by
	Wiener filtering and that to be maximally robust, the watermark should
	have a power spectrum matching the one of the original signal.

	The problem of robust embedding and extraction based on spread
	spectrum is revisited here in light of a more general model of the
	cover signal and of the watermark channel. Most of the approaches
	introduced so far consider that the cover signal can be modelled as an
	ergodic wide sense stationary Gaussian random process. This assumption
	is rarely satisfied for real signals. We assume instead that it can be
	modelled as the realization of a set of independent non identically
	distributed Gaussian random variables (referred to as non i.i.d.
	signals). The attack channel is considered to be of the type amplitude
	scaling and additive white Gaussian noise
	(SAWGN)~\cite{EggBaumlGirod2002}. The game-theoretic resolution of the
	problem with weighted distortion measures leads to a characterization
	of optimum attack domains. By maximizing the watermarking channel
	signal to noise ratio, we then derive a closed-form expression of the
	watermark spectral density corresponding to the best defense. The
	performance limits of the approach in terms of hiding capacity are
	then analyzed. The approach can be seen as a generalization of
	previous work to the case of non i.i.d. Gaussian sources, considering
	weighted distortion measures and a more general SAWGN attack channel,
	with the exhibition of closed-form expressions for a practical
	embedding and extraction scheme.

\section{PROBLEM STATEMENT}
	Let ${\underline {\mathbf b}} = \{b_1, b_2, \ldots , b_n\}$ with $b_i
	\in \{-1, +1\}$ $\forall i \in \{1, 2, \ldots , n\}$ be the message
	to be embedded in a host signal ${\underline {\mathbf x}}$. Many
	approaches introduced so far assume that the signal ${\underline
	{\mathbf x}}$ can be modelled as an ergodic zero-mean wide sense
	stationary Gaussian random
	process~\cite{MoulinSullivan2000,SuEggersGirod2001}. This assumption
	is rarely satisfied for real signals or for content adaptive
	watermarks. We assume instead that the host signal ${\underline
	{\mathbf x}}$ can be modelled as the realization of a set of non
	stationary Gaussian random variables ${\underline {\mathbf X}}=\{X_1,
	X_2, \ldots , X_m\}$ where $X_i \sim {\mathcal N}(0, \sigma_{X_i})$.
	The information is then used as a key for indexing pseudo-random noise
	sequences which are additively combined with the signal. Let
	${\underline {\mathbf G}}$ be a $n \times m$ matrix composed of $n$
	pseudo-random generated vectors ${\underline {\mathbf G_j}} \in \{-1,
	+1\}^m$. The watermarked signal is obtained by 
	\vspace{-0.2cm} \begin{equation}
		\label{eq:mark}
		y_i = x_i + w_i = x_i + \alpha_i \sum_{j=1}^n G_{i,j} b_j \text{,}
	\vspace{-0.2cm} \end{equation}
	where $x_i$ represents the $i^{\text{th}}$ site of the host signal and
	$y_i$ the corresponding watermarked site. In order to extract each
	embedded bit $b_i$, a correlation product between the vector
	${\underline {\mathbf G_j}}$ and ${\underline {\mathbf y}}$ is
	computed. The term $\alpha_i$ is a weighting factor allowing to adjust
	the amplitude (or energy) of the mark. In the following we derive a
	closed-form expression of this parameter in the case of SAWGN attacks,
	with weighted distortion measures and non i.i.d Gaussian cover signals.
	The attack channel is often assumed to be
	AWGN~\cite{piva:detection98}. This model assumes that the
	distortion induced by the attack is independent of the
	watermarked signal, hence can hardly apply to attacks such as
	filtering and compression. More accurate models assuming that
	the distortion depends on the watermarked signal and based on
	linear filtering plus additive noise have been considered
	in~\cite{SuEggersGirod2001,MoulinIvanovic2001}. Here, we
	consider that the attacked signal ${\underline {\mathbf y'}}$
	can be expressed as
	\vspace{-0.2cm} \begin{equation}
		\label{eq:attack}
		y'_i = \gamma_i y_i + \delta_i 
		= \gamma_i x_i + \gamma_i \alpha_i \sum_{j=1}^n
		G_{i,j} b_j + \delta_i 
		\text{,}
	\vspace{-0.2cm} \end{equation}
	where $\gamma_i$ is an attenuation factor on each watermarked
	site. This amounts to consider the attack channel as a SAWGN
	channel (amplitude scaling by the factor $\gamma_i$, and
	additive white Gaussian noise of $\delta_i \sim {\mathcal
	N}(0, \sigma_{\delta_i}^2)$).

	The distortion measure is defined as a weighted sum of the MSE
	on each sample of the host signal, in order to reflect the
	perceptual quality. The embedding distortion is therefore
	given by 
	\vspace{-0.2cm} \begin{equation}
		D_{xy} = E \left[
			\sum_{i=1}^m \varphi_i^2 \left( y_i - x_i
			\right)^2 
		\right]
		= \sum_{i=1}^m \varphi_i^2 n \alpha_i^2 
		\text{,}
	\vspace{-0.2cm} \end{equation}
	where $\varphi_i$ is a perceptual factor. Similarly, the
	expected attack distortion is given by
	\vspace{-0.2cm} \begin{equation}
		\label{eq:dxyp}
		D_{xy'} = \sum_{i=1}^m \varphi_i^2 \left(
			\sigma_{X_i}^2 \left( 
				1-\gamma_i 
			\right)^2 + n \gamma_i^2 \alpha_i^2 +
			\sigma_{\delta_i}^2 
		\right) \text{.}
	\vspace{-0.2cm} \end{equation}

\section{MAP WATERMARK ESTIMATION}
	The {\it maximun a posteriori} (MAP) estimation of the bit
	$b_j$ is defined as 
	\vspace{-0.2cm} \begin{equation}
		\label{estimator}
		\widehat{b_j} = \arg \max_{b_j} \left\{
			P(B_j=b_j \, | \, {Y'}^m = {\underline
			{\mathbf y'}}) 
		\right\} \text{.}
	\vspace{-0.2cm} \end{equation}
	The a posteriori probability ${\mathcal P} = P(B_j=b_j
	\, | \, {Y'}^m = {\underline {\mathbf y'}})$ can be rewritten
	(using Bayes law) as 
	\vspace{-0.15cm} \begin{equation}
		\label{eq:proba}
		{\mathcal P} =  \frac{
			P({Y'}^m = {\underline {\mathbf y'}} \, | \,
			B_j=b_j) \times P(B_j=b_j)
		}{
			P({Y'}^m = {\underline {\mathbf y'}})
		} \text{.}
	\vspace{-0.2cm} \end{equation}
	Since the received vector ${\underline {\mathbf y'}}$ is
	fixed, and that no a priori knowledge on the message
	${\underline {\mathbf b}}$ is assumed, we have the a posteriori
	probability ${\mathcal P} \propto P({Y'}^m={\underline
	{\mathbf y'}} \, | \, B_j=b_j)$, where $\propto$ denotes an
	obvious renormalization. Assuming that the watermarked sites
	are independent, it can be
	shown~\cite{pateuxleguelvouitguillemot01} that the quantity
	can be expressed as a product of Gaussian distributions of the
	form ${\mathcal P}_i \sim {\mathcal N}(0, \gamma_i^2
	(\sigma_{X_i}^2 + \alpha_i^2 (n-1)) + \sigma_{\delta_i}^2)$,
	i.e. as 
	\vspace{-0.2cm} \begin{eqnarray}
		\label{eq11}
		\mathcal{P} &\propto& \prod_{i=1}^m
		\frac{
			1
		}{
			\sqrt{2\pi} \sqrt{
				\gamma_i \sigma_{X_i^2} + \alpha_i^2 (n-1)) +
				\sigma_{\delta_i}^2
			}
		} \nonumber \\
		&& \exp \left[
			-\frac{
				\left(
					y'_i - \gamma_i \alpha_i
					b_j G_{i,j}
				\right)^2
			}{
				2 \left(
					\gamma_i^2 (\sigma_{X_i}^2 + \alpha_i^2 (n-1)) + \sigma_{\delta_i}^2
				\right)
			}
		\right] \text{,} \\
		&\propto& \frac{
			C
		}{
			2
		} \exp \frac{ 
			- \Lambda 
		}{
			2
		} 
		\label{eq:gaussianChannel1} \text{,}
	\vspace{-0.2cm} \end{eqnarray}
	where $C$ is a constant and
	\vspace{-0.2cm} \begin{equation}
		\Lambda = \sum_{j=1}^n \frac{ 
			\left(
				b_j - \widehat{b_j}
			\right)^2
		}{
			\sigma_{{b_j}}^2
		} 
		\label{eq:gaussianChannel2}
	\vspace{-0.2cm} \end{equation}
	with
	\vspace{-0.2cm} \begin{eqnarray}
		\label{estimator2}
		\widehat{b_j} &=& \frac{ 
			\sum_{i=1}^m \frac{
				\gamma_i \alpha_i y'_i G_{i,j}
			}{	
				\gamma_i^2 (\sigma_{X_i}^2 + \alpha_i^2 (n-1)) + \sigma_{\delta_i}^2
			}
		}{
			\sum_{i=1}^m \frac{ 
				\gamma_i^2 \alpha_i^2 
			}{
				\gamma_i^2 (\sigma_{X_i}^2 + \alpha_i^2 (n-1)) + \sigma_{\delta_i}^2
			}
		} \label{eq:moyenne} \text{,} \\
		\sigma_{{b_j}}^{2} &=& \left( 
			\sum_{i=1}^m \frac{ 
				\gamma_i^2 \alpha_i^2
			}{
				\gamma_i^2 (\sigma_{X_i}^2 + \alpha_i^2 (n-1)) + \sigma_{\delta_i}^2
			}
		\right)^{-1} \text{.}
		\label{eq:variance}
	\vspace{-0.2cm} \end{eqnarray}
	The term $\widehat{b_j}$ represents the optimal estimator.
	From Eqn.(\ref{eq:gaussianChannel1})
	and~(\ref{eq:gaussianChannel2}), watermarking channel can be
	seen as a gaussian channel. Estimator's performance can be
	measured in terms of the signal to noise ratio $E_b/N_0$ of
	the watermarking channel. This quantity is defined as the
	ratio between the energy of the embedded bit and the overall
	noise introduced by the cover signal ($\sigma_{X_i}^2$), by
	the other embedded bits ($\alpha_i^2(n-1)$) and by the attack
	(i.e. $\sigma_{\delta_i}^2$).  It is then expressed as 
	\vspace{-0.2cm} \begin{equation}
		\label{eq:perf}
		\frac{E_b}{N_0} =  \frac{
			E(b_j^2)
		}{
			\sigma_{b_j}^2 
		} = \frac{
			1
		}{
			\sigma_{b_j}^2
		} = \sum_{i=1}^m \rho_i \text{,}
	\vspace{-0.2cm} \end{equation}
	where
	\vspace{-0.2cm} \begin{equation}
		\label{eq:omega}
		\rho_i = \frac{
			\alpha_i^2 \gamma_i^2
		}{
			\gamma_i^2 (\sigma_{X_i}^2 + \alpha_i^2 (n-1)) + \sigma_{\delta_i}^2
		} \text{.}
	\vspace{-0.2cm} \end{equation}

\section{GAME-THEORETIC RESOLUTION}
\begin{figure}[b]
		\centerline{\psfig{figure=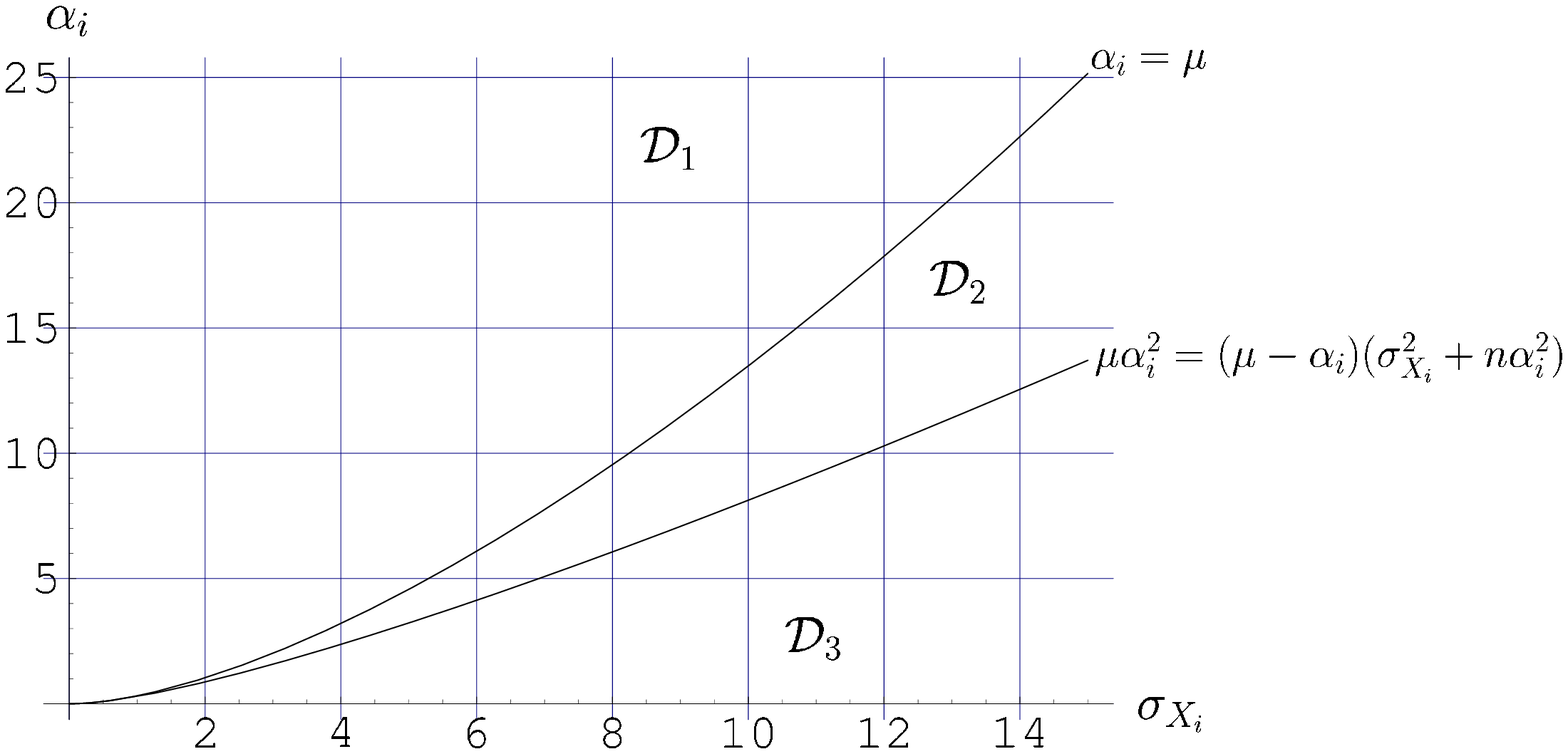,width=6cm}}
		\vspace{-2mm}
		\caption{Domains defined by the validity constraints $\gamma_i^{\star} \geq 0$ and
		$\sigma_{\delta_i^{\star}} \geq 0$ ($\lambda=0.2$,
		$\chi=0.002$, $n=1$ bit).}
		\label{fig:regions}
		\vspace{-0.5cm}
	\end{figure}

	The optimization of the embedding and attack parameters can be
	formulated as a game between an attacker and a hider. The attack
	searches the two vectors $\underline {\mathbf {\gamma}}$ and
	$\underline{\mathbf {\sigma}_{\delta}}$ minimizing the extractor
	performance (i.e. $E_b/N_0$) while maintaining the distortion below
	an acceptable level ($D_{xy'} < D_{xy'}^{\max}$). This problem
	can be solved by a Lagrangian optimization: 
	\vspace{-0.2cm} \begin{equation}
		\nonumber 
		\left(
			\underline {\mathbf \gamma^{\star}}, \underline{\mathbf \sigma_{\delta^{\star}}}
		\right) = \arg \min_{
			\underline {\mathbf {\gamma}},
			\underline {\mathbf \sigma_{\delta}}
		} \left\{
			J_{\lambda} = \frac{E_b}{N_0} +
			\lambda \left[
				D_{xy'} - D_{xy'}^{\max}
			\right]
		\right\} \text{,}
	\vspace{-0.2cm} \end{equation}
	where $\lambda>0$ is a Lagrangian multiplier. From
	Eqn.(\ref{eq:dxyp}) and~(\ref{eq:perf}) it appears that
	$J_\lambda$ is an additive functional. The optimization can
	then be made separately on each $J_{\lambda, i}$ given by 
	\vspace{-0.2cm} \begin{equation}
		J_{\lambda, i}(\gamma_i, \sigma_{\delta_i}) = \rho_i +
		\lambda \varphi_i^2
		\left( 
			\sigma_{X_i}^2 (1-\gamma_i)^2 + n \gamma_i^2 \alpha_i^2 +
			\sigma_{\delta_i}^2
		\right)
		\nonumber
	\vspace{-0.2cm} \end{equation}
	by setting its derivatives with respect to the attack
	parameters $\gamma_i$ and $\sigma_{\delta_i}$
	\vspace{-0.2cm} \begin{eqnarray}
		\frac{ 
			\partial J_{\lambda, i} 
		}{ 
			\partial \gamma_i
		} &=& 2 \frac{ 
			\gamma_i \alpha_i^2 \sigma_{\delta_i}^2 
		}{ 
			\left( 
				\varP \right)^2
		} \nonumber \\
		&&+ 2 \lambda \varphi_i^2
		\left(
			n \gamma_i \alpha_i^2 - \sigma_{X_i}^2
			\left(
				1-\gamma_i
			\right)
		\right) \text{,} 
	\vspace{-0.2cm} \end{eqnarray}
	and
	\vspace{-0.2cm} \begin{eqnarray}
		\label{eq:derivb} 
		\frac{
			\partial J_{\lambda, i}
		}{
			\partial {\sigma_{\delta_i}}
		} = \frac{
			-2 \gamma_i^2 \alpha_i^2 \sigma_{\delta_i}
		}{
			\left(
				\varP
			\right)^2
		} + 2 \lambda \varphi_i^2 \sigma_{\delta_i}
	\vspace{-0.2cm} \end{eqnarray}
	to zero on the validity domain. The resolution of the
	resulting set of equations leads to the following two expressions for
	$\sigma_{\delta_i}^2$:
	\vspace{-0.2cm} \begin{equation}
		\label{delta1}
		\sigma_{\delta_i}^2 = - \gamma_i \left[
			n \gamma_i \alpha_i^2 - \sigma_{X_i}^2
			\left(
				1-\gamma_i
			\right)
		\right]
	\vspace{-0.2cm} \end{equation}
	and
	\vspace{-0.2cm} \begin{equation}
		\label{delta2}
		\sigma_{\delta_i}^2 = \gamma_i
		\left[
			\frac{
				\alpha_i
			}{
				\sqrt{\lambda} \varphi_i
			} - \gamma_i
			\left(
				\sigma_{X_i}^2 + \alpha_i^2 (n-1)
			\right)
		\right] \text{.}
	\vspace{-0.2cm} \end{equation}
	Equating Eqn.(\ref{delta1}) and~(\ref{delta2}) leads to the
	optimum values 
	\vspace{-0.2cm} \begin{equation}
		\label{eq:gamma3}
		\gamma_i^{\star} = \frac{
			\sqrt{\lambda} \varphi_i \sigma_{X_i}^2
			- \alpha_i
		}{
			\sqrt{\lambda} \varphi_i
			\alpha_i^2
		}
	\vspace{-0.2cm} \end{equation}
	and
	\vspace{-0.2cm} \begin{equation}
		\label{optimdelta}
		\sigma_{\delta_i^{\star}}^2 = \gamma_i^{\star} \left( 
			{\gamma_w}_i - \gamma_i^{\star}
		\right) \left(
			\sigma_{X_i}^2 +n \alpha_i^2
		\right) \text{,}
	\vspace{-0.2cm} \end{equation}
	where ${\gamma_w}_i = \frac{\sigma_{X_i}^2}{\sigma_{X_i}^2 +
	\sigma_{w_i}^2}$, and $\sigma_{w_i}^2 = n\alpha_i^2$. The term
	${\gamma_w}_i$ represents the response of a Wiener filter.
	Since the attack parameters $\sigma_{\delta_i^{\star}}^2$ and
	$\gamma_i^{\star}$ must verify $\sigma_{\delta_i^{\star}}^2 \geq 0$ and
	$\gamma_i^{\star} \geq 0$, the solutions of Eqn.(\ref{eq:gamma3})
	and~(\ref{optimdelta}) are valid only for 
	\vspace{-0.2cm} \begin{eqnarray}
		\label{eq:constraint1}
		\mu - \alpha_i &\geq& 0 \\
		\label{eq:constraint2}
		\left(
			\alpha_i - \mu
		\right) \left(
			\sigma_{X_i}^2 + n \alpha_i^2
		\right) + \mu \alpha_i^2 &\geq& 0 \text{,}
	\vspace{-0.2cm} \end{eqnarray}
	where $\mu = \sqrt{\lambda} \varphi_i \sigma_{X_i}^2$. This
 	set of inequations defines three domains ${\cal D}_1$, ${\cal
 	D}_2$ and ${\cal D}_3$ shown in Fig.~\ref{fig:regions}.

	The optimum attacks can then be characterized in terms of the
	domains of validity of the attack parameters $\gamma_i$ and
	$\sigma_{\delta_i}$. Let us first consider their limits of
	validity ($\gamma_i=0$ and $\sigma_{\delta_i}=0$). If
	$\gamma_i^{\star} = 0$ (the marked value $y_i$ is erased), a minimum
	is obtained for $\sigma_{\delta_i^{\star}} = 0$. A greater value for
	the additive noise will increase $D_{xy'}$ but will not
	decrease $E_b/N_0$.  This attack is referred to as the {\bf
	Erase} attack. If $\sigma_{\delta_i}=0$, another minimum is
	given by $\gamma_i^{\star} = {\gamma_w}_i$. This attack is a {\bf
	Wiener} filtering. The last attack (defined for $\gamma_i^{\star} >
	0$ and for $\sigma_{\delta_i^{\star}} > 0$) is a combination of
	filtering and additive Gaussian noise. This is called here the
	{\bf Intermediate} attack. Table~\ref{tab:jlambda} gives the
	corresponding expressions of the cost function $J_{\lambda,
	i}(\gamma_i, \sigma_{\delta_i})$ denoted $J_E$, $J_W$ and
	$J_I$ for respectively the erase, Wiener and intermediate attacks.
	To find the optimum attack, one has to find on each domain
	(defined in terms of $\alpha_i$ and $\sigma_{X_i}$), the
	attack that will minimize $J_{\lambda, i}(\gamma_i,
	\sigma_{\delta_i})$. From table~\ref{tab:jlambda} and
	constraints~(\ref{eq:constraint1})-(\ref{eq:constraint2}) the
	minimum values of $J_{\lambda, i}(\gamma_i,
	\sigma_{\delta_i})$, in the domains of validity of $\gamma_i$
	and $\sigma_{\delta_i}$, are given by $J_E$ and $J_W$ on
	${\cal D}_1$ and ${\cal D}_3$ respectively
	(see~\cite{pateuxleguelvouitguillemot01} for details).
	Similarly, on ${\cal D}_2$, $J_I \leq J_E$ and $J_I \leq J_W$.
	Thus, if the validity constraint $\gamma_i^{\star} \geq 0$ of the
	Intermediate attack domain is satisfied, the optimum attack is
	given by the Intermediate attack (with parameters given by
	Eqn.(\ref{eq:gamma3}) and~(\ref{optimdelta})). Otherwise, the
	attacker should use instead the Erase or the Wiener solution.

	\begin{table}
		\begin{center}
		\begin{tabular}{c|c}
			& Value of $J_{\lambda, i}(\gamma_i, \sigma_{\delta_i})$ \\
			\hline
			$J_E$ & $\lambda \varphi_i^2 \sigma_{X_i}^2$ \\
			$J_W$ & $\frac{
				\alpha_i^2
			}{
				\sigma_{X_i}^2 + \alpha_i^2
				\left(
					n-1
				\right)
			} + \lambda \frac{
				n \varphi_i^2
				\alpha_i^2 \sigma_{X_i}^2
			}{
				\sigma_{X_i}^2 + n
				\alpha_i^2
			}$ \\
			$J_I$ & $2 \sqrt{\lambda} \varphi_i
			\frac{
				\sigma_{X_i}^2
			}{
				\alpha_i
			} -1 + \lambda \varphi_i^2 \sigma_{X_i}^2
			\left(
				1- \frac{
					\sigma_{X_i}^2
				}{
					\alpha_i^2
				}
			\right)$ \\
		\end{tabular}
		\vspace{-2mm}
		\caption{Cost function $J_{\lambda, i}(\gamma_i, \sigma_{\delta_i})$ for the different types of attack (Erase, Wiener, Intermediate).\label{tab:jlambda}}
                \end{center}
		\vspace{-0.8cm}
        \end{table}

	Given the optimum attack, we then search the parameters
 	$\alpha_i$ (strength of the watermark) that maximize
 	$E_b/N_0$, under constraints of a maximum watermarked signal
 	distortion ($D_{xy}^{\max}$). This leads to a Lagrangian approach:
	\vspace{-0.2cm} \begin{equation}
		{\underline {\mathbf \alpha^{\star}}} = \arg
		\max_{\underline {\mathbf \alpha}} 
		\left\{
			J_\chi = J_\lambda - \chi 
			\left[
				D_{xy} - D_{xy}^{\max}
			\right]
		\right\} \text{,}
	\vspace{-0.2cm} \end{equation}
	where $\chi>0$ is a Lagrangian multiplier. The cost function
 	$J_\chi$ being the additive functional $J_\chi = \sum_i^m
 	J_{\lambda,i} (\gamma_i, \sigma_{\delta_i}) - \chi \left[
 	D_{xy} - D_{xy}^{\max} \right]$, the optimization can be
 	carried out separately on each $J_{\chi, i}(\alpha_i) =
 	J_{\lambda, i}(\gamma_i, \sigma_{\delta_i}) - \chi n
 	\varphi_i^2 \alpha_i^2$. Let us consider the three attack
 	strategies. In the case of the Erase attack, i.e. $(\alpha_i,
 	\sigma_{X_i}) \in {\cal D}_1$, $J_{\chi, i}(\alpha_i) =
 	\lambda \varphi_i^2 \sigma_{X_i}^2 - \chi n \varphi_i^2
 	\alpha_i^2$. The function $J_{\chi, i}(\alpha_i)$ is a
 	decreasing function and the minimum valid value of $\alpha_i$
 	is given by $\alpha_i^{\star} = \sqrt{\lambda} \varphi_i
 	\sigma_{X_i}^2$. In ${\cal D}_2$, setting the derivative 
	\vspace{-0.2cm} \begin{equation}
		\label{eq:derivD2}
		\frac{
			\partial J_{\chi, i}
		}{
			\partial \alpha_i
		} = - 2 \frac{
			\sqrt{\lambda} \varphi_i \sigma_{X_i}^2
		}{
			\alpha_i^2
		} + 2 \frac{
			\lambda \varphi_i^2 \sigma_{X_i}^4
		}{
			\alpha_i^3
		}  - 2 \chi n \varphi_i^2 \alpha_i
	\vspace{-0.2cm} \end{equation}
	to zero leads to $\mu^2 - \mu \alpha_i - \chi n \varphi_i^2
	\alpha_i^4=0$ where $\mu=\sqrt{\lambda} \varphi_i
	\sigma_{X_i}^2$. The derivative is negative for $\alpha_i =
	\mu$ and positive for $\alpha_i = 0$. The polynomial being
	monotonous on the interval $[0; \mu]$, one can conclude that
	Eqn.(\ref{eq:derivD2}) has a valid solution on $[0; \mu]$. If
	the derivative is negative on ${\cal D}_2$, the solution
	adopted is $\alpha_i$ such that $\mu \alpha_i^2 = (\mu -
	\alpha_i)(\sigma_{X_i}^2 + n \alpha_i^2)$. Let $n \alpha_i^2 =
	\sigma_{W_i}^2$, $n\lambda = \lambda'$ and $n\chi = \chi'$.
	Let us assume that $\sigma_{W_i}^2$ is very close to
	$\alpha_i^2(n-1)$. In the Wiener case ($(\gamma_i,
	\sigma_{\delta_i}) \in {\cal D}_3$), the cost function
	$J_{\chi, i}(\alpha_i)$ expressed in terms of $\sigma_{W_i}^2
	= n \alpha_i^2$ is given by 
	\vspace{-0.2cm} \begin{equation}
		J_{\chi, i}(\sigma_{W_i}^2) = \left(
			1+ \lambda' \varphi_i^2 \sigma_{X_i}^2
		\right) \frac{
			\sigma_{W_i}^2
		}{
			\sigma_{W_i}^2 + \sigma_{X_i}^2
		} - \chi' \varphi_i^2 \sigma_{W_i}^2 \text{.}
	\vspace{-0.2cm} \end{equation}
	Setting the derivative of $J_{\chi, i}(\sigma_{W_i}^2)$ with
	respect to $\sigma_{W_i}^2$
	\vspace{-0.2cm} \begin{equation}
		\frac{
			\partial J_{\chi, i}
		}{
			\partial \sigma_{W_i}^2
		} = \left(
			1+ \lambda' \varphi_i^2 \sigma_{X_i}^2
		\right) \frac{
			\sigma_{X_i}^2
		}{
			\left(
				\sigma_{W_i}^2 + \sigma_{X_i}^2
			\right)^2
		} - \chi' \varphi_i^2
	\vspace{-0.2cm} \end{equation}
	to zero, given that $\sigma_{W_i}^2 = n \alpha_i^2$, leads to
	\vspace{-0.2cm} 

\begin{figure}[t]
		\centerline{\psfig{figure=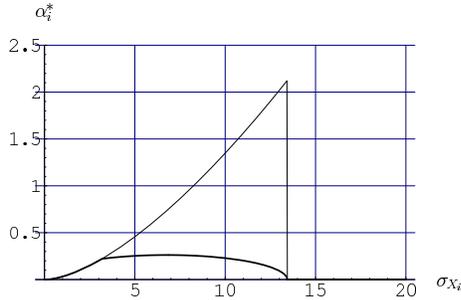,width=6cm}}
		\vspace{-2mm}
		\caption{Optimum values of $\alpha_i$ for
		$\lambda=0.002$, $\chi=0.0028$ and $n=100$ bits.}
		\label{fig:alpha}
		\vspace{-0.5cm}
	\end{figure}

	\vspace{-0.2cm} \begin{equation}
		\alpha_i^{\star} = \sqrt{
			\frac{
				\sigma_{X_i} \sqrt{
					1+ \lambda' \varphi_i^2 \sigma_{X_i}^2
				} - \sqrt{\chi'} \varphi_i
				\sigma_{X_i}^2
			}{
				\sqrt{\chi'} n \varphi_i
			}
		} \text{.}
	\vspace{-0.2cm} \end{equation}
	This provides a closed-form of the optimum embedding parameter
 	$\alpha_i$ in terms of the host signal power spectrum
 	($\sigma_{X_i}^2$) and for an SAWGN attack. The Wiener
 	filtering can restore the signal, hence may lead to $D_{xy'} <
 	D_{xy}$. This can be avoided by filtering the watermarked
 	signal (after embedding). The distortion measure is then
 	$D_{xy} = \sum_{i=1}^m \varphi_i^2 \frac{ \sigma_{X_i}^2
 	\sigma_{W_i}^2 }{\sigma_{X_i}^2 + \sigma_{W_i}^2 }$, where
 	$\sigma_{W_i}^2 = n \alpha_i^2$. The resolution of the problem
 	leads to new parameters
 	(see~\cite{pateuxleguelvouitguillemot01} for details):
 	$\alpha_i^{\star} \simeq \sqrt{\lambda} \varphi_i  \sigma_{X_i}^2$.
 	Fig.~\ref{fig:alpha} illustrates the variations of the
 	parameter $\alpha_i$ in terms of $\sigma_{X_i}^2$ for both
 	approaches i.e., without (bold curve) and with a Wiener
	post-filtering (light curve) of the watermarked signal (with
	$\varphi_i=(1+\sigma_{X_i})^{-1/2}$).
	Unlike~\cite{CoxKilianLeightonShamoon97}, it can be observed
	that for high values of $\sigma_{X_i}^2$, no watermark
	can be robustly embedded on the corresponding sites.

\section{RESULTS}
\begin{figure}
		\centerline{\psfig{figure=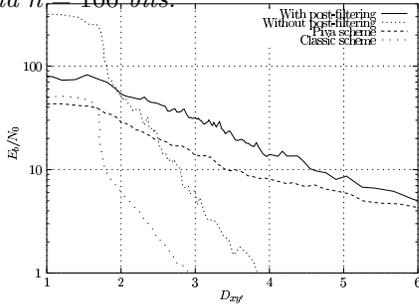,width=5.5cm}}
		\vspace{-2mm}
		\caption{Performances of the proposed scheme in presence of a compression attack JPEG.}
		\label{fig:compa_jpeg}
		\vspace{-0.5cm}
	\end{figure}

	The approach has been tested on images against techniques
	using $\alpha_i=constant$, and $\alpha_i=c
	|x_i|$~\cite{piva:detection98}, considering embedding in the
	wavelet transform domain. Fig.~\ref{fig:compa_jpeg} depicts
	the respective $E_b/N_0$ performances in terms of the attack
	distortion. A message of $156$ bits is embedded (i.e. $n=156$)
	in the $512 \times 512$ gray scale {\it Lena} image (i.e.
	$m=262~144$). The Lagragian multipliers $\lambda$ and $\chi$
	are set so that $D_{xy}=D_{xy}^{\text{max}}$ and
	$D_{xy'}=D_{xy'}^{\text{max}}$. The embedding parameters have been tuned in
	order to get the same perceptual distortion $D_{xy}/m=1$
	(with $\varphi_i=(1+\sigma_{X_i})^{-1/2}$) with the different
	techniques. The watermarked image has been attacked with a
	lossy compression JPEG, from 95\% to 5\% quality. The tests,
	using the Stirmark benchmark have shown that the technique is
	robust to all the non-geometric attacks. 

\section{CONCLUSION}
	This paper provides an information-theoretic analysis of information
	hiding in non i.i.d signals with perceptual distortion metrics.
	Note that previous work, when considering perceptual watermaking
	was often led by intuition. Here we have derived closed-form expressions
	of the different extraction and embedding parameters, revealing an
	efficient and practical information hiding system.

\small
\bibliographystyle{plain}
\bibliography{eusipco}

\end{document}